\documentclass[11pt,english]{article}
\usepackage{ae,aecompl}
\usepackage[T1]{fontenc}
\usepackage[latin9]{inputenc}
\usepackage{geometry}
\usepackage{array}
\usepackage{amsmath}
\usepackage{amsthm}
\usepackage{amssymb}
\usepackage{graphicx}
\usepackage{setspace}
\usepackage[authoryear]{natbib}
\makeatletter

\providecommand{\tabularnewline}{\\}

\theoremstyle{plain}
\newtheorem{thm}{\protect\theoremname}
\theoremstyle{plain}
\newtheorem{prop}[thm]{\protect\propositionname}

\usepackage{setspace}

\usepackage[bottom]{footmisc}
\renewcommand{\hat}{\widehat}
\renewcommand{\tilde}{\widetilde}

\DeclareSymbolFont{extrasymbols}{OMS}{cmsy}{m}{n}
\DeclareMathDelimiter{\lVert}
  {\mathopen}{extrasymbols}{"6B}{largesymbols}{"0D}
\DeclareMathDelimiter{\rVert}
  {\mathclose}{extrasymbols}{"6B}{largesymbols}{"0D}

\usepackage{xcolor}
\definecolor{mycolor}{HTML}{0072BB}

\let\oldprod\prod
\renewcommand\prod{\textstyle\oldprod}

\makeatother

\usepackage{babel}
\providecommand{\propositionname}{Proposition}
\providecommand{\theoremname}{Theorem}

\begin{document}

\date{November 24, 2025}

\title{Pre-Training Estimators for Structural Models: Application to Consumer
Search\vspace{2em}}
\author{Yanhao 'Max' Wei and Zhenling Jiang\thanks{Max: Marshall School of Business, University of Southern California.
Zhenling: The Wharton School, University of Pennsylvania. Emails:
yanhaowe@usc.edu and zhenling@wharton.upenn.edu. We thank the comments
from Eric Bradlow, Yufeng Huang, Yewon Kim, Guillaume Pouliot, Navdeep
Sahni, Stephan Seiler, Raphael Thomadsen, Song Yao, Yuyan Wang, Raluca
Ursu, Matthijs Wildenbeest, Dennis Zhang, and the participants at
the Chicago marketing seminar, Stanford marketing seminar, WashU marketing
seminar, HongKong PolyU marketing seminar, European Quant Marketing
Seminar, Consumer Search and Switching Costs Workshop, and Conference
on Frontiers in Machine Learning and Economics (Philly Fed). }}
\maketitle

\begin{abstract}
\begin{singlespace}We develop pre-trained estimators for structural
econometric models. The estimator uses a neural net to recognize the
structural model's parameter from data patterns. Once trained, the
estimator can be shared and applied to different datasets at negligible
cost and effort. Under sufficient training, the estimator converges
to the Bayesian posterior given the data patterns. As an illustration,
we construct a pretrained estimator for a sequential search model
(available at \texttt{pnnehome.github.io}). Estimation takes only
seconds and achieves high accuracy on 12 real datasets. More broadly,
pretrained estimators can make structural models much easier to use
and more accessible.\end{singlespace}\vspace{2em}

\newpage
\end{abstract}

\section{Introduction}

Pre-trained machine learning models, ranging from popular image recognition
models to large language models, are achieving significant success
(e.g., AlexNet, GPT). These pretrained models allow end users to greatly
reduce the costs of pattern-recognition tasks. Take image recognition
for example: instead of collecting images, labeling them, and training
a neural net from scratch, users can directly apply a pretrained neural
net to recognize objects in their images.

This paper explores the use of pretraining in structural estimation.
We develop pretrained estimators for structural econometric models.
The bulk of computational cost and researcher effort occur only once
during the estimator's construction. Subsequently, the estimator can
be shared and applied to different datasets with negligible cost and
effort. More specifically, a pretrained estimator assumes a given
structural model. We use the structural model to simulate datasets,
which are then used to train a neural net to recognize the structural
model's parameter from a set of data patterns (similar to how neural
nets recognize objects from images). With sufficient training, the
estimator converges to the Bayesian posterior of the parameter given
the data patterns. We refer to the estimator as the pretrained neural
net estimator (hereafter, pretrained NNE).

Pretrained NNE makes structural models easier to use and more accessible.
In practice, estimating a structural model is often technically complex
and time-consuming. Yet, the value of structural models -- such as
uncovering micro-founded preferences and enabling counterfactuals
-- is realized only after estimation. A pretrained NNE can be built
once and then shared, allowing its users to easily and quickly estimate
a structural model across applications. For users, the cost and effort
of estimation become comparable to those in reduced-form regressions.
Further, pretrained estimators can broaden the scope of applications
of structural models. For example, one may apply pretrained estimators
continuously to real-time data, enabling the integration of structural
models into real-time algorithms (e.g., dynamic pricing, recommenders,
bandit experiments). Another benefit of pretrained NNE is that it
helps preserve privacy in estimation because it relies only on aggregate
data patterns as input.

As an illustration, this paper focuses on sequential search models.
These models characterize consumer search and purchase choices, and
are finding growing application especially with online shopping data.
Search models are estimated by simulated maximum likelihood (SMLE),
but estimation is challenging. The vast number of search and purchase
combinations makes it infeasible to directly simulate likelihood.
To this end, the literature has introduced several workarounds over
time, such as likelihood smoothing or specialized simulation algorithms
(see \citealt{ursu2023sequential}). Even then, a single estimation
run is expensive, let alone the multiple runs needed to tune optimizers,
try starting points, tune smoothing factors, tune simulation algorithms,
etc. Importantly, these tunings have to be performed separately for
each application. More broadly, similar estimation challenges exist
in many other structural models.

Pretrained NNE offers an alternative way to estimate a structural
model. It does so by learning a direct mapping from data patterns
to the structural model's parameter. Conceptually, we can think of
a pretrained NNE as follows. Research papers using structural estimation
often start with reduced-form evidence before estimating the structural
model. Although reduced-form results do not directly translate into
an estimate of the structural model's parameter, they offer useful
clues. For example, if customers are more likely to search lower-price
items, we expect the price coefficient in the search model to be more
negative. If customers make more searches over weekends, we know weekend
customers have either a low search cost or a low outside utility.
More generally, we can imagine that with a rich set of such reduced-form
patterns, an expert familiar with the search model can guesstimate
the search model parameter value. The present paper's exercise is
to leverage machine learning to create such an ``expert.''

\begin{figure}
\begin{centering}
\includegraphics[scale=0.5]{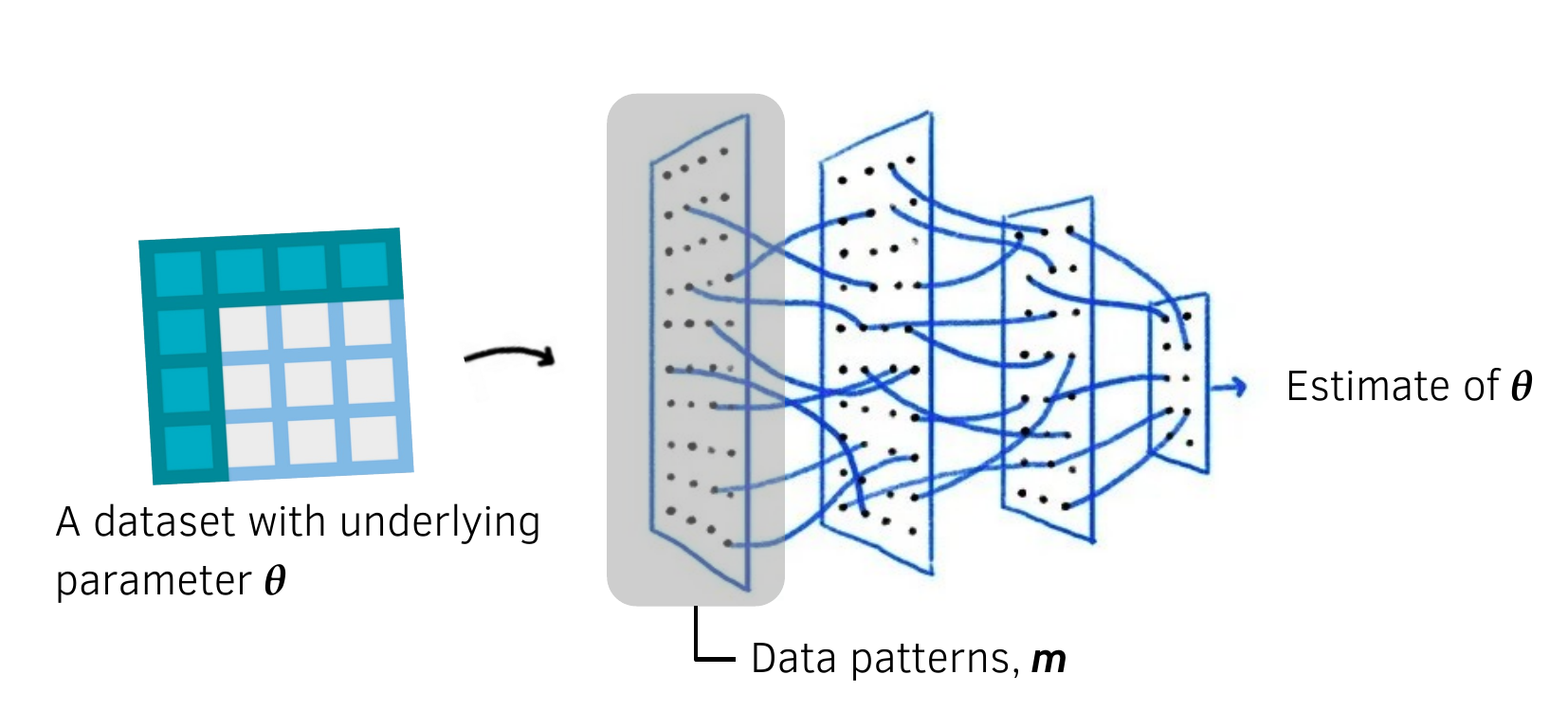}
\par\end{centering}
\caption{An Illustration of NNE\protect\label{fig:illustrate}}

\end{figure}

We describe a standard sequential search model in Section \ref{sec:search}.
Then, in Section \ref{sec:pnne} we describe how to pretrain an NNE
for this search model. At a high level, the training uses a large
number of datasets simulated from the search model. Let $\boldsymbol{\theta}$
denote the search model parameter. For each draw of $\boldsymbol{\theta}$
from a prior, we can use the search model to simulate a dataset ${\cal D}$.
Use $\boldsymbol{m}$ to collect some data patterns of ${\cal D}$
(e.g., summary statistics, regression coefficients). Repeating this
process for $L$ draws of $\boldsymbol{\theta}$ gives us $\{\boldsymbol{\theta}^{(\ell)},\boldsymbol{m}^{(\ell)}\}_{\ell=1}^{L}$
(we currently use $L\simeq$ 1 million). These $L$ pairs serve as
the training examples for a neural net to learn a mapping from $\boldsymbol{m}$
to $\boldsymbol{\theta}$. The trained neural net becomes the aforementioned
``expert'' capable of estimating the search model. Figure \ref{fig:illustrate}
offers an illustration of this neural net.

From a theoretical perspective, the estimate of NNE reflects a Bayesian
posterior. Under mild conditions, as $L\rightarrow\infty$, the pretrained
NNE converges to the posterior mean of $\boldsymbol{\theta}$ given
$\boldsymbol{m}$. This mean is known as a limited-information posterior
mean, and it is one's best estimate for the structural model's parameter
after seeing the data patterns.

We provide our pretrained NNE at: \texttt{pnnehome.github.}\texttt{io}.
It can be applied off-the-shelf to estimate the search model: users
only need to supply data, no further tuning is required, and the computation
typically takes seconds on a laptop. Specifically, users supply a
dataset ${\cal D}$ consisting of two components: (i) $\boldsymbol{y}$,
which includes search and purchase choices, and (ii) $\boldsymbol{x}$,
which includes product attributes that affect utility, consumer attributes
that affect outside utility, and advertising attributes that affect
search cost. The current version of our pretrained NNE can accommodate
datasets with $n\geq1000$ consumers, 2 to 8 product attributes, 0
to 5 consumer attributes, and 0 to 2 advertising attributes. These
ranges are configurable before pretraining.

Section \ref{sec:applications} evaluates the pretrained NNE using
12 real datasets. We conduct two sets of tests. The first set consists
of Monte Carlo studies that estimate the search model on real-data
attributes but simulated consumer choices under a ``true'' $\boldsymbol{\theta}$.
Pretrained NNE is faster than SMLE by 3-4 orders of magnitude. Further,
it recovers $\boldsymbol{\theta}$ with accuracy equal to or higher
than SMLE. The result holds even when compared against the SMLE that
uses a specialized simulation algorithm for search models that takes
considerable effort to implement (\citealt{jiang2021consumer}). The
second set of tests estimates the search model entirely from real
data. Because the true $\boldsymbol{\theta}$ is unknown, we compare
the pretrained NNE's estimates with SMLE's estimates. Their estimates
turn out to be similar. The pretrained NNE is again faster by 3-4
orders of magnitude.

An important detail in pretraining NNE is to specify the data patterns
$\boldsymbol{m}$ --- a representation of the input ${\cal D}$.
In computer vision or language processing, representations of inputs
(images or texts) typically must be learned via training. By contrast,
in structural models, economic theory and intuition often provide
sufficient guidance for us to specify an $\boldsymbol{m}$ that can
identify $\boldsymbol{\theta}$. This greatly reduces the training
complexity. To optimize estimation efficiency, alternative specifications
of $\boldsymbol{m}$ can be compared using the neural net's accuracy
on validation examples (i.e., a held-out subset of training examples).
Importantly, this effort to select $\boldsymbol{m}$ arises only in
pretraining and is not borne by the users of a pretrained NNE. For
our search model, we choose $\boldsymbol{m}$ to include the coefficients
from a set of regressions that describe consumer search and purchase
behaviors.\footnote{In \citet{wei2024estimating}, $\boldsymbol{m}$ includes only data
moments. Expanding $\boldsymbol{m}$ to include reduced-form regression
coefficients leads to a substantial improvement in estimation accuracy
for the search model.}

This paper extends the standard (non-pretrained) NNE in \citet{wei2024estimating}.
In a standard NNE, all training examples assume the same observed
attributes $\boldsymbol{x}$ from a real dataset. Thus, the trained
net is applicable only to that dataset. In contrast, a pretrained
NNE is applicable across different datasets. To achieve this, in each
training example we draw $\boldsymbol{x}$ from a distribution (and
then use the search model to simulate $\boldsymbol{y}$ given the
drawn $\boldsymbol{x}$). This distribution serves the role of a prior
for $\boldsymbol{x}$. We specify it to reflect attributes common
in real applications, covering diverse scenarios that vary in the
number of attributes, marginal distributions, correlations, and within-
versus cross-consumer variation. We use a large $L$ to cover these
scenarios well. It is important to note that an NNE (standard or pretrained)
still assumes a specific structural model. Therefore, pretrained NNE
is more suited for established structural models that are poised for
broader applications, whereas standard NNE is more suited for newly
proposed or ad hoc structural models.

Finally, we emphasize that this paper's goal is to explore the feasibility
of pretrained estimators for structural models, rather than to develop
a full-fledged tool for estimating search models. For the latter task,
there remains room for improvement, such as allowing richer unobserved
consumer heterogeneity and incorporating instruments to address possible
endogeneity. These extensions are non-trivial and we discuss them
in Section \ref{sec:conclusion}.

\subsection{Literature\protect\label{subsec:literature}}

This paper is related to several streams of research. First, this
paper joins a recent literature that seeks to leverage machine learning
for econometric estimation. In particular, the literature has studied
the use of flexible machine learning models to better capture individual-level
heterogeneity (e.g., \citealt{wager2018estimation}, \citealt{athey2019generalized},
\citealt{farrell2020deep}), to better incorporate high-dimensional
nuisance parameters (e.g., \citealt{chernozhukov2018double}), and
to construct instrumental variables (e.g., \citealt{hartford2017deepiv},
\citealt{lewis2018adversarial}, \citealt{singh2020machine}). A main
theme of this literature so far is to use machine learning to flexibly
describe the relations between variables in data. The current paper
differs in that the relations between data variables are described
by an existing econometric model (e.g., sequential search), and we
leverage machine learning to estimate the parameter of this econometric
model. Recent works sharing this theme include \citet{wei2024estimating}
and \citet{kaji2023adversarial}. In particular, \citet{kaji2023adversarial}
borrow ideas from adversarial learning, and propose training a discriminator
to construct the objective function for estimating the econometric
model. This approach is conceptually different from ours.

Second, we build on the successful idea of pretraining in machine
learning practice. Pretraining involves training a machine learning
model to learn general capabilities which can then be applied to various
specific tasks. In natural language processing, pretrained models
like BERT and GPT demonstrate remarkable capabilities in many language
tasks (\citealt{devlin2018bert}, \citealt{brown2020gpt3}). In computer
vision, pretrained neural nets such as AlexNet and ResNet have been
applied widely (\citealt{krizhevsky2012imagenet}, \citealt{he2016res}).
A review can be found in \citet{han2021pre}. Pretrained models have
considerably improved the productivity of researchers and practitioners.
The current paper aims to bring the benefits of pretraining to structural
estimation in economics and marketing. Pretrained estimators are ``Train
Once, Deploy Anywhere.'' Their availability will make structural
estimation more accessible to researchers and practitioners.

Third, this paper adds to the literature that seeks to enable model
estimation while preserving data privacy. Data privacy is becoming
increasingly relevant to marketers. Various methods have been proposed
to preserve privacy in estimation, such as noise infusion (\citealt{abowd2012dynamically},
\citealt{avella2021privacy}), generating synthetic data (\citealt{schneider2018flexible},
\citealt{anand2023using}), and estimation using aggregate data (e.g.,
\citealt{musalem2009bayesian}). The present paper advances privacy-preserving
estimation using aggregate data. Pretrained NNE requires only reduced-form
data patterns to estimate a structural model. Sharing reduced-form
data patterns, which are aggregated across individuals, poses far
less privacy concern for data owners than sharing individual-level
data.

\section{Consumer search model\protect\label{sec:search}}

In this section, we describe a standard (sequential) search model
and the typical data used to estimate it. The next section will present
the pretrained NNE for this search model.

\subsection{The model\protect\label{subsec:search-model}}

Sequential search models describe how consumers search and choose
from a set of available options. One example where these models are
increasingly applied is online browsing and shopping. In this context,
consumer is given a webpage with a list of products; she decides which
products to search (i.e., click) and which product to buy. A survey
of sequential search models can be found in \citet{ursu2023sequential}.

The current paper focuses on a fairly standard search model. Formally,
there are $n$ consumers (or search sessions) and $J$ products per
consumer. Let vector $\boldsymbol{x}_{ij}^{(prod)}$ collect the attributes
of product $j$ for consumer $i$. The utility of consumer $i$ from
product $j$ is:
\[
u_{ij}=\boldsymbol{\beta}'\boldsymbol{x}_{ij}^{(prod)}+\mu_{ij}+\varepsilon_{ij}.
\]
There are two shocks, $\mu_{ij}$ and $\varepsilon_{ij}$. The consumer
observes $\mu_{ij}$ pre-search but observes $\varepsilon_{ij}$ only
post-search. Both shocks are distributed as $N(0,1)$. While normalizing
one shock is without loss of generality, normalizing both shocks is
not. Nevertheless, many applications have normalized both shocks because
it can be empirically difficult to identify the variances of both
shocks. We follow this practice. See \citet{ursu2023sequential} for
more discussion.

Let vector $\boldsymbol{x}_{i}^{(cons)}$ collect the attributes of
consumer $i$ (e.g., demographics, purchase history). These attributes
affect the utility of an outside option as follows. The shock $\varepsilon_{i0}$
is known pre-search and follows $N(0,1)$.
\[
u_{i0}=\eta_{0}+\boldsymbol{\eta}'\boldsymbol{x}_{i}^{(cons)}+\varepsilon_{i0}.
\]

Each consumer has one free search. Afterwards, a search cost is incurred
for each additional search. We allow for some product-level attributes
that affect search cost but not utility, collected in $\boldsymbol{x}_{ij}^{(ads)}$.
We will refer to them as advertising or marketing attributes. One
example of such attributes is the product rankings on the webpage.
The search cost $c_{ij}$ is specified as
\[
\log(c_{ij})=\alpha_{0}+\boldsymbol{\alpha}'\boldsymbol{x}_{ij}^{(ads)}.
\]

We use $d^{(prod)}$, $d^{(ads)}$, and $d^{(cons)}$ to denote the
numbers of attributes in $\boldsymbol{x}_{ij}^{(prod)}$, $\boldsymbol{x}_{ij}^{(ads)}$
, and $\boldsymbol{x}_{i}^{(cons)}$, respectively. Let $d\equiv d^{(prod)}+d^{(ads)}+d^{(cons)}$
denote the number of all attributes. Let $\boldsymbol{\theta}=(\boldsymbol{\beta}',\boldsymbol{\eta}',\boldsymbol{\alpha}',\eta_{0},\alpha_{0})'$
denote our search model's parameter. The dimension of $\boldsymbol{\theta}$
is $d+2$. 

The search model characterizes: (i) which products the consumer will
search, (ii) which products the consumer will buy, and (iii) the order
in which the consumer searches the products. We shall not delve into
deriving the optimal choices of the consumer. It suffices to note
that these choices are typically computed using the optimal decision
rule in \citet{weitzman1979optimal}.

It is useful to note that the search model can also be applied to
contexts outside online browsing and purchase. For example, geo-location
data are increasingly available. They allow one to study consumers'
search behaviors in offline shopping. Shop visits and subsequent purchases
can be modeled by a sequential search model. An interesting aspect
is that the distances from shops to elevators or entrances may affect
search costs (and thus be considered as a marketing attribute).

\subsection{The data\protect\label{subsec:search-data}}

Next, we describe the typical data used to estimate the search model.
In doing so, we also make clear the range of data to which our pretrained
NNE is applicable.

Continuing with the notation so far, we use $i$ to index consumers
and $j$ to index products. Let $\boldsymbol{x}_{ij}^{(prod)}$ collect
product attributes, $\boldsymbol{x}_{ij}^{(ads)}$ collect advertising
attributes, and $\boldsymbol{x}_{i}^{(cons)}$ collect consumer attributes.
A dataset ${\cal D}$ can be written as
\begin{align}
{\cal D} & =\{\boldsymbol{x},\;\boldsymbol{y}\},\label{eq:dataset}\\
\text{where}: & \quad\boldsymbol{x}=\{\{\boldsymbol{x}_{ij}^{(prod)},\boldsymbol{x}_{ij}^{(ads)}\}_{j=1}^{J},\:\boldsymbol{x}_{i}^{(cons)}\}_{i=1}^{n},\nonumber \\
 & \quad\boldsymbol{y}=\{\{\boldsymbol{y}_{ij}\}_{j=1}^{J}\}_{i=1}^{n}.\nonumber 
\end{align}
Here, vector $\boldsymbol{y}_{ij}$ collects two outcome variables:
\begin{itemize}
\item $y_{ij}^{(search)}$ is a dummy indicating whether the option $j$
is searched;
\item $y_{ij}^{(buy)}$ is a dummy indicating whether $j$ is bought.
\end{itemize}
In some applications, the order of searches is also observable, so
$\boldsymbol{y}_{ij}$ can have a third element $y_{ij}^{(order)}$.
However, because the search order is not always observable, in this
paper we choose not to make use of search order information (so $y_{ij}^{(order)}$
is not required). Conceptually, it is straightforward to pretrain
a separate NNE that uses the search order information.

Our pretrained NNE assumes that $\boldsymbol{x}_{ij}^{(prod)}$, $\boldsymbol{x}_{ij}^{(ads)}$
and $\boldsymbol{x}_{i}^{(cons)}$ are all standardized to have zero
means and unit variances. The means and variances here are computed
across $i$ and $j$, e.g., the mean for $\boldsymbol{x}_{ij}^{(prod)}$
is $\frac{1}{nJ}\sum_{i=1}^{n}\sum_{j=1}^{J}\boldsymbol{x}_{ij}^{(prod)}$.
If the attributes are not standardized, we standardize them before
using the pretrained NNE to estimate $\boldsymbol{\theta}$. We then
rescale the estimate of $\boldsymbol{\theta}$ to make it consistent
with respect to the original, non-standardized attributes.

\begin{table}
\caption{Specified Ranges for Dataset Size\protect\label{tab:data-dim}}

\smallskip{}

\centering{}%
\begin{tabular}{cc>{\centering}p{0.05\paperwidth}>{\centering}p{0.05\paperwidth}>{\centering}p{0.05\paperwidth}>{\centering}p{0.05\paperwidth}>{\centering}p{0.18\paperwidth}c}
\hline 
 &  & $d^{(prod)}$ & $d^{(ads)}$ & $d^{(cons)}$ & $J$ & $n$ & \tabularnewline
\hline 
min &  & 2 & 0 & 0 & 15 & 1e3{\small$\begin{array}{c}
\text{}\\
\\\end{array}$} & \tabularnewline
max &  & 8 & 2 & 5 & 35 & {\small$\begin{array}{c}
\infty\;\text{ (application)}\\
2e4\text{ (pretraining)}
\end{array}$} & \tabularnewline
\hline 
\end{tabular}
\end{table}

The size of ${\cal D}$ is determined by the numbers of attributes
($d^{(prod)}$, $d^{(ads)}$, $d^{(cons)}$), number of products per
consumer ($J$), and number of consumers ($n$). These numbers vary
across applications. Our current version of pretrained NNE can be
applied to datasets of sizes within the ranges specified in Table
\ref{tab:data-dim}. These ranges are configurable before pretraining.

A note about Table \ref{tab:data-dim} is that no cap is imposed on
$n$ for \textit{applying} the pretrained NNE. But a cap on $n$ is
necessary in\textit{ pretraining}, because we cannot simulate data
of infinite size. This cap is currently set at $2e4$. The pretrained
NNE is not directly applicable to data larger than this cap. However,
in practice, we can split a very large dataset and apply the pretrained
NNE to each split. We then take the average of the estimates across
the splits.

It is difficult to estimate $\boldsymbol{\theta}$ if the data contain
too few searches or purchases. Therefore, we do not pretrain NNE to
accommodate such data. Specifically, we look at:
\begin{itemize}
\item Buy rate: fraction of consumers that bought some product (not outside
good).
\item Search rate: fraction of consumers that did at least one non-free
search.
\end{itemize}
We require that the buy rate is at least 0.5\% and the search rate
is at least 1\%. Again, these thresholds are configurable prior to
pretraining. During pretraining, simulated datasets where search rates
or buy rates are lower than the thresholds are dropped.

As we will describe later, the pretraining incorporates non-normally
distributed attributes. Nevertheless, the pretrained NNE may still
perform poorly if certain attributes are too heavily skewed or have
too many outliers. Therefore, we recommend using the usual practices
to prepare data as in any empirical analysis. These include trimming,
winsorizing, transforming variables, and dropping any attribute that
causes (almost) collinearity.

\section{Pretrained NNE\protect\label{sec:pnne}}

This section describes the construction of pretrained NNE. Although
we aim to make the description self-contained, more emphasis is given
to where the pretrained NNE differs from the standard NNE (\citealt{wei2024estimating}). 

\subsection{Overview\protect\label{subsec:pnne-overview}}

Let ${\cal D}$ denote a generic dataset as defined in expression
(\ref{eq:dataset}). Let $\boldsymbol{\theta}$ be the parameter vector
of the search model specified in Section \ref{subsec:search-model}.
The task of NNE is to map ${\cal D}$ to an estimate of $\boldsymbol{\theta}$.
It does so through two functions, $\boldsymbol{f}$ and $\boldsymbol{h}$.
\[
\hat{\boldsymbol{\theta}}=\boldsymbol{f}\big[\:\underbrace{\boldsymbol{h}({\cal D})}_{\boldsymbol{m}}\:\big].
\]

\begin{itemize}
\item The first function $\boldsymbol{h}$ calculates a set of data patterns
of ${\cal D}$, collected in vector $\boldsymbol{m}=\boldsymbol{h}({\cal D})$.
This function is specified by us (rather than trained via examples).
\item The second function $\boldsymbol{f}$ is a neural net that maps from
the data patterns to an estimate of $\boldsymbol{\theta}$. We write
$\hat{\boldsymbol{\theta}}=\boldsymbol{f}(\boldsymbol{m})$.
\end{itemize}
Conceptually, one could forgo specifying $\boldsymbol{h}$ and instead
train a neural net mapping directly from ${\cal D}$ to $\boldsymbol{\theta}$.
But the size of ${\cal D}$ is much larger than $\boldsymbol{m}$.
In addition, the mapping from ${\cal D}$ to $\boldsymbol{\theta}$
tends to be much more complex than from $\boldsymbol{m}$ to $\boldsymbol{\theta}$.
So the specification of function $\boldsymbol{h}$ substantially reduces
the complexity of the learning task. Further, the use of aggregate
data patterns also allows privacy-preserving estimation, as discussed
in Section \ref{subsec:literature}.

In general, the vector $\boldsymbol{m}=\boldsymbol{h}({\cal D})$
may include any data pattern of ${\cal D}$. Ideally, we want to include
data patterns that are most informative about $\boldsymbol{\theta}$
and yet inexpensive to compute. The foremost candidates are various
data moments (hence the notation $\boldsymbol{m}$), such as means,
variances, correlations. Further candidates are coefficients of reduced-form
regressions. The idea of using reduced-form coefficients as ``moments''
originated from indirect inference (\citealt{gourieroux1993indirect}).
Including reduced-form coefficients in $\boldsymbol{m}$ turns out
to substantially improve the estimation accuracy of NNE, and is an
important extension to \citet{wei2024estimating}.

The detailed specification of $\boldsymbol{h}$ in our pretrained
NNE is given in the appendix. Below, we give a summary.
\begin{itemize}
\item First included in $\boldsymbol{m}$ are data size and summary statistics:
$d^{(prod)}$, $d^{(ads)}$, $d^{(cons)}$, $n$, $J$, buy rate,
search rate, the average number of searches per consumer.
\item To capture search decisions, we include the coefficients from a logit
regression that predicts $y_{ij}^{(search)}$. Similarly, to capture
purchase decisions, we include the coefficients from a multinomial
logit regression that predicts $y_{ij}^{(buy)}$ conditional on search,
i.e., $y_{ij}^{(search)}=1$. These regressions are at the product
level.
\item We include the coefficients from a set of regressions that predict
outcomes at the consumer level (e.g., whether consumer $i$ bought
anything, whether $i$ made any non-free search).
\end{itemize}
Next, we turn to the function $\boldsymbol{f}$, which plays a central
role in the pretrained NNE. The function is a neural net trained using
simulated datasets. The training is conducted as follows.
\begin{itemize}
\item We use the search model to simulate $L$ datasets, ${\cal D}^{(1)}$,
${\cal D}^{(2)}$, ..., ${\cal D}^{(L)}$, each under a different
parameter vector, $\boldsymbol{\theta}^{(1)}$, $\boldsymbol{\theta}^{(2)}$,
..., $\boldsymbol{\theta}^{(L)}$.
\item For each $\ell=1,2,...,L$, we compute the data patterns $\boldsymbol{m}^{(\ell)}=\boldsymbol{h}({\cal D}^{(\ell)})$. 
\item Each pair $\{\boldsymbol{m}^{(\ell)},\boldsymbol{\theta}^{(\ell)}\}$
is a training example. We use $\{\boldsymbol{m}^{(\ell)},\boldsymbol{\theta}^{(\ell)}\}_{\ell=1}^{L}$
to train a neural net to predict $\boldsymbol{\theta}$ from $\boldsymbol{m}$.
\end{itemize}
We have left out three details in this training process: (i) the configuration
of the neural net, (ii) the distribution from which we draw $\boldsymbol{\theta}^{(\ell)}$,
and (iii) the distribution from which we draw the product, consumer,
and advertising attributes in ${\cal D}^{(\ell)}$. These details
are discussed in Section \ref{subsec:pnne-details}. Generally speaking,
the distributions in (ii) and (iii) should be broad enough for NNE
to accommodate a variety of real datasets in applications, but not
so broad that NNE wastes resources learning scenarios that are unlikely
in reality. A good balance makes the training more effective for a
given $L$. Meanwhile, it is important for us to point out the pretrained
NNE's estimates in applications (Section \ref{sec:applications})
are not sensitive to the specifications of (i)-(iii).

The description so far has focused on point estimates. For the standard
error of $\hat{\boldsymbol{\theta}}$, we recommend bootstrapping.
This boils down to applying the pretrained NNE to resampled copies
of ${\cal D}$. Because applying the pretrained NNE to each copy of
${\cal D}$ incurs very little cost, bootstrapping standard errors
is not computationally expensive at all.\footnote{The bootstrapping here omits the uncertainty in $\boldsymbol{f}$
resulting from the neural net training. However, given the large number
of training examples ($L\simeq1$ million), this uncertainty tends
to be small relative to the sampling uncertainty of ${\cal D}$.}

Finally, a practical issue is how NNE behaves when the search model
is misspecified for the dataset in a real application. NNE takes data
patterns as input and outputs $\hat{\boldsymbol{\theta}}$. This output
is given so that, when the search model is simulated at $\hat{\boldsymbol{\theta}}$,
the resulting data patterns are close to the input data patterns.
Importantly, this behavior does not require the input data patterns
to be generated by the search model. In this sense, NNE is well-behaved
under misspecification.\footnote{Also see Section \ref{subsec:pnne-interpret} that shows the pretrained
NNE converges to $\mathbb{E}(\boldsymbol{\theta}|\boldsymbol{m})$,
which tends to be well-behaved under misspecification provided that
the observed $\boldsymbol{m}$ is within the model-implied support
of $\boldsymbol{m}$.} However, this behavior does require the input data patterns to have
positive probability under the search model and priors. If not, the
input will lie outside the range of inputs seen in training and the
neural net has to extrapolate. We call such cases ``ill-suited''
for the pretrained NNE, and it is useful to detect them in practice.
We provide a detection method in the appendix.

\subsection{Connection to Bayesian estimation\protect\label{subsec:pnne-interpret}}

The eventual test for a pretrained model is to see how it performs
in applications. We do so for the pretrained NNE in Section \ref{sec:applications}
and find that it provides accurate and sensible estimates. That said,
from a theoretical standpoint, it is important to understand the econometric
meaning of the pretrained NNE's estimate. Below, we will explain that
pretrained NNE has a familiar Bayesian interpretation, in that it
calculates the posterior mean of $\boldsymbol{\theta}$ given $\boldsymbol{m}$.
In words, it provides the best estimate of the structural model's
parameter $\boldsymbol{\theta}$ given the data patterns $\boldsymbol{m}$.

Continuing with the notation in (\ref{eq:dataset}), let ${\cal D}=\{\boldsymbol{x},\boldsymbol{y}\}$
denote a dataset. In particular, $\boldsymbol{x}$ and $\boldsymbol{y}$
collect the observations \textit{across all consumers}. The vector
$\boldsymbol{m}=\boldsymbol{h}({\cal D})$ denotes a set of data patterns.
We start with a result with the standard (non-pretrained) NNE, studied
in \citet{wei2024estimating}. The main difference between standard
NNE and pretrained NNE is that the former is trained for a specific
real dataset. The training examples do not simulate $\boldsymbol{x}$
but directly use the observed $\boldsymbol{x}$ from that real dataset. 

The standard NNE converges to $\mathbb{E}(\boldsymbol{\theta}|\boldsymbol{m},\boldsymbol{x})$
as $L\rightarrow\infty$\textit{ }(see \citealt{wei2024estimating}).
This result does not require large $n$ and holds for any value of
$n$. Also note that we make it explicit that the expectation conditions
on a specific $\boldsymbol{x}$.\footnote{The notation of \citet{wei2024estimating} omits the conditioning
on $\boldsymbol{x}$ because the paper assumes an underlying probability
space that is already conditional on the observed $\boldsymbol{x}$.} The $\mathbb{E}(\boldsymbol{\theta}|\boldsymbol{m},\boldsymbol{x})$
is known as a limited-information Bayesian posterior mean (e.g., \citealt{kim2002limited}),
where ``limited-information'' indicates that the posterior relies
on less information than the full data. To compare, the full-information
posterior is $\mathbb{E}(\boldsymbol{\theta}|\boldsymbol{y},\boldsymbol{x})$.
Ideally, we would like to know the full-information posterior, but
when computing it is difficult, the limited-information posterior
provides a practical alternative.

We now turn to the pretrained NNE. In this case, the training examples
do not assume a specific $\boldsymbol{x}$, but instead draw $\boldsymbol{x}$
from a distribution. Consequently, the estimate by the pretrained
NNE converges to $\mathbb{E}(\boldsymbol{\theta}|\boldsymbol{m})$
as $L\rightarrow\infty$, where the expectation no longer conditions
on $\boldsymbol{x}$.

We provide a formal statement for this convergence result. To be precise,
we consider the joint probability distribution between $\boldsymbol{\theta}$
and $\boldsymbol{m}$, $P(\boldsymbol{\theta},\boldsymbol{m})$. This
joint distribution is implied by the search model, our specification
of $\boldsymbol{m}$, and the distributions from which we draw $\boldsymbol{\theta}$
and $\boldsymbol{x}$ in pretraining. Specifically, in terms of probability
densities, 
\begin{equation}
p(\boldsymbol{\theta},\boldsymbol{m})=\int p(\boldsymbol{m}|\boldsymbol{\theta},\boldsymbol{x})p(\boldsymbol{\theta})p(\boldsymbol{x})\:d\boldsymbol{x}.\label{eq:distribution}
\end{equation}
Here, $p(\boldsymbol{m}|\boldsymbol{\theta},\boldsymbol{x})$ is the
likelihood of observing data patterns $\boldsymbol{m}$ under the
search model. Terms $p(\boldsymbol{\theta})$ and $p(\boldsymbol{x})$
are densities from which we draw $\boldsymbol{\theta}$ and $\boldsymbol{x}$
in pretraining, and $p(\boldsymbol{\theta},\boldsymbol{x})=p(\boldsymbol{\theta})p(\boldsymbol{x})$
as we will draw them independently. With the probability density in
equation (\ref{eq:distribution}), we have
\[
\mathbb{E}(\boldsymbol{\theta}|\boldsymbol{m})=\frac{\int\boldsymbol{\theta}p(\boldsymbol{\theta},\boldsymbol{m})d\boldsymbol{\theta}}{\int p(\boldsymbol{\theta},\boldsymbol{m})d\boldsymbol{\theta}}.
\]
Each training example $\{\boldsymbol{\theta}^{(\ell)},\boldsymbol{m}^{(\ell)}\}$
is a draw from $P(\boldsymbol{\theta},\boldsymbol{m})$. We use $\boldsymbol{f}_{L}$
to denote the neural net trained with $L$ training examples. That
is, $\boldsymbol{f}_{L}=\min_{\boldsymbol{\zeta}\in{\cal F}_{L}}\sum_{\ell=1}^{L}\lVert\boldsymbol{\zeta}(\boldsymbol{m}^{(\ell)})-\boldsymbol{\theta}^{(\ell)}\rVert^{2}$,
where ${\cal F}_{L}$ is the same class of single-hidden-layer neural
nets defined in \citet{wei2024estimating}. The result below states
that the function $\boldsymbol{f}_{L}$ converges to $\mathbb{E}(\boldsymbol{\theta}|\boldsymbol{m})$
as a function of $\boldsymbol{m}$.
\begin{prop}
\label{prop:convergence}Suppose: (i) $\Theta$ is compact, (ii) $\boldsymbol{m}$
has a compact support, (iii) $\mathbb{E}(\boldsymbol{\theta}|\boldsymbol{m})$
is a continuous function of $\boldsymbol{m}$. We have $\int\lVert\boldsymbol{f}_{L}(\boldsymbol{m})-\mathbb{E}(\boldsymbol{\theta}|\boldsymbol{m})\rVert^{2}dP(\boldsymbol{m})\rightarrow0$
in probability as $L\rightarrow\infty$.
\end{prop}

The continuity condition should hold in most structural models as
it only requires the posterior mean not to change abruptly for a slight
change in the observed data patterns. In comparison, the compactness
conditions can be more restrictive. While the compactness of $\Theta$
is common in econometric settings, it technically precludes unbounded
priors such as the normal distribution. In practice, we can trim the
prior at extreme values to ensure a compact $\Theta$. Similarly,
the support-compactness of $\boldsymbol{m}$ can be satisfied if we
trim extreme data patterns in training. Nevertheless, in applications
(Section \ref{sec:applications}), we find that the pretrained NNE's
estimates show very little difference whether trimming is used in
pretraining. Finally, it is worth emphasizing that the conditions
in Proposition \ref{prop:convergence} are sufficient but not necessary.

Given this convergence result, a natural question is how to interpret
$\mathbb{E}(\boldsymbol{\theta}|\boldsymbol{m})$. Formally, it is
another limited-information posterior mean. However, for a more intuitive
perspective, consider a practical scenario where a researcher wants
to apply the search model to a company's data. For privacy reasons,
the company provides the researcher with only aggregate data patterns
$\boldsymbol{m}$, instead of the full data $\{\boldsymbol{x},\boldsymbol{y}\}$.
In this case, the researcher's best estimate for the search model
parameter will be precisely $\mathbb{E}(\boldsymbol{\theta}|\boldsymbol{m})$.
From equation (\ref{eq:distribution}), we see that for $\mathbb{E}(\boldsymbol{\theta}|\boldsymbol{m})$
to be well defined, the researcher should form a prior belief over
$\boldsymbol{\theta}$, $p(\boldsymbol{\theta})$, and a prior belief
over $\boldsymbol{x}$, $p(\boldsymbol{x})$. In the pretrained NNE,
these priors are the distributions from which $\boldsymbol{\theta}$
and $\boldsymbol{x}$ are drawn during pretraining (details in Section
\ref{subsec:pnne-details}).

\subsection{Pretraining details\protect\label{subsec:pnne-details}}

Section \ref{subsec:pnne-overview} gave an overview of our pretrained
NNE, leaving out several details. We provide these details below.
Readers interested mainly in the applications may skip these details
without much loss of continuity.

\subsubsection{The neural net}

A neural net accepts fixed-length input and gives fixed-length output.
But the search model parameter $\boldsymbol{\theta}$ has a size of
$d+2$, where $d$ is the total number of attributes and it varies
across different datasets. We address this issue through padding.
Specifically, recall that $d\equiv d^{(prod)}+d^{(ads)}+d^{(cons)}$.
Use subscript ``max'' to denote the maximum allowed value. For example,
Table \ref{tab:data-dim} sets $d_{max}^{(prod)}=8$ and $d_{max}=8+2+5=15$.
We configure the neural net to always output $d_{max}+2$ values.
Out of these output values, only $d+2$ values are useful. The rest
are the padded ``idle'' values. Using ellipses to represent idle
values, we can write the output as:
\[
(\enspace\alpha_{0}\enspace,\enspace\boldsymbol{\alpha}'\enspace,\underbrace{\enspace\cdot\cdot\cdot\enspace}_{d_{max}^{(ads)}-d^{(ads)}},\enspace\eta_{0}\enspace,\enspace\boldsymbol{\eta}'\enspace,\underbrace{\enspace\cdot\cdot\cdot\enspace}_{d_{max}^{(cons)}-d^{(cons)}},\enspace\boldsymbol{\beta}'\enspace,\underbrace{\enspace\cdot\cdot\cdot\enspace}_{d_{max}^{(prod)}-d^{(prod)}}).
\]
During training, the idle output values are ignored by the loss function.
In applications, the idle output values are discarded.

A similar arrangement applies to the neural net's input. The length
of $\boldsymbol{m}$ varies by the numbers of attributes in the data.
For our current settings in Table \ref{tab:data-dim}, the maximal
length of $\boldsymbol{m}$ is 164. Any shorter $\boldsymbol{m}$
is padded with zeros to this maximal length before we feed it into
the neural net. The exact detail of $\boldsymbol{m}$ is given in
the appendix.

We currently use $L\simeq$ 1e6 training examples (after dropping
``corner'' datasets with very low search or buy rates; see Section
\ref{subsec:search-data}). We hold out $5e4$ examples for validation.
Increasing $L$ further only marginally reduces the validation loss.
We also use the hold-out validation to choose the neural net architecture.
While Proposition \ref{prop:convergence} is formally stated with
single-hidden-layer neural nets, in practice we find that using a
few more hidden layers can offer lower validation losses. We end up
using three hidden layers with 512, 256, and 256 nodes. The pretrained
NNE's estimates in applications (Section \ref{sec:applications})
are not sensitive to the neural net architecture.

\subsubsection{The parameter prior}

When generating the training datasets, we draw $\boldsymbol{\theta}^{(1)}$,
$\boldsymbol{\theta}^{(2)}$, ..., $\boldsymbol{\theta}^{(L)}$ from
a distribution of $\boldsymbol{\theta}$. As explained in Section
\ref{subsec:pnne-interpret}, this distribution of $\boldsymbol{\theta}$
effectively plays the role of a prior for $\boldsymbol{\theta}$.
We want the prior to be sufficiently broad to cover possible values
of $\boldsymbol{\theta}$ in real applications, but not so broad that
many training datasets become unrealistic and thus wasteful. Below,
we explain the prior used in our pretrained NNE. We have also tried
several alternative priors. The pretrained NNE's estimates in applications
(Section \ref{sec:applications}) are not sensitive to the prior specification.

Unlike Bayesian estimation on a single dataset, a unique challenge
in specifying our prior is that we need to accommodate datasets with
varying dimensions ($d^{(prod)}$, $d^{(cons)}$, $d^{(ads)}$). Consider
key statistics such as buy rate and search rate. In real applications,
there is no inherent reason for us to expect these key statistics
to vary systematically across datasets of different dimensions. Therefore,
across training datasets, we would like the key statistics not to
vary systematically by data dimensions, either.

This desire rules out common priors such as $\beta_{k}\sim N(0,1)$
for $k=1,...,d^{(prod)}$. With this prior, the buy rates would be
systematically higher in training datasets with larger $d^{(prod)}$.
To see why, note this prior implies $\lVert\boldsymbol{\beta}\rVert^{2}\sim\chi_{d^{(prod)}}^{2}$
which has an expectation equal to $d^{(prod)}$. Consider a simple
case where $\boldsymbol{x}_{ij}^{(prod)}\sim N(\boldsymbol{0},\boldsymbol{I})$.
Then, the variance of $\boldsymbol{\beta}'\boldsymbol{x}_{ij}^{(prod)}$
for any given $\boldsymbol{\beta}$ equals $\lVert\boldsymbol{\beta}\rVert^{2}$,
which in expectation increases with $d^{(prod)}$. A larger variance
makes it more likely for at least one product to be better than outside
option, leading to a higher buy rate.

To this end, we specify a prior where $\lVert\boldsymbol{\beta}\rVert$
is invariant to $d^{(prod)}$: 
\begin{equation}
\boldsymbol{\beta}\sim\sigma\cdot\chi_{1}\cdot\boldsymbol{\nu},\label{eq:prior}
\end{equation}
where $\sigma$ is a constant, $\chi_{1}$ is chi distributed, and
$\boldsymbol{\nu}$ is a unit vector with a uniform distribution on
the unit sphere. An implication of this prior is $\lVert\boldsymbol{\beta}\rVert^{2}\sim\sigma^{2}\cdot\chi_{1}^{2}$,
which is invariant to $d^{(prod)}$. We use the same type of prior
for $\boldsymbol{\eta}$ and $\boldsymbol{\alpha}$.

Intuitively, the prior (\ref{eq:prior}) can also be understood as
letting the overall impact of product attributes (as measured by $\lVert\boldsymbol{\beta}\rVert$)
be invariant to $d^{(prod)}$. This means that when there are more
product attributes, we expect that each individual attribute has a
lower chance to have a large effect on utility. This belief seems
to align with what we see in real applications, where typically only
a few ``key'' product attributes have substantial effects on consumers'
search and purchase decisions.

The discussion so far has left out the priors for the intercepts,
$\alpha_{0}$ and $\eta_{0}$, which directly affect search cost and
outside utility. We use normal priors for these intercepts, with their
means and variances specified to produce realistic search rates and
buy rates in training datasets (neither too small nor too close to
1).

\subsubsection{Attributes ($\boldsymbol{x}$)}

The generation of training examples requires us to specify a distribution
of $\boldsymbol{x}$. Specifically, to generate a training dataset
${\cal D}^{(\ell)}$, we first draw $\boldsymbol{x}^{(\ell)}$ from
this distribution. Then, given $\boldsymbol{x}^{(\ell)}$ and $\boldsymbol{\theta}^{(\ell)}$,
we use the search model to simulate $\boldsymbol{y}^{(\ell)}$, and
let ${\cal D}^{(\ell)}=\{\boldsymbol{x}^{(\ell)},\boldsymbol{y}^{(\ell)}\}$.
Conceptually, the distribution of $\boldsymbol{x}$ should be broad
enough to cover common values of $\boldsymbol{x}$ in real applications,
but not so broad that many training datasets become unrealistic and
thus wasteful. Below, we describe the distribution of $\boldsymbol{x}$
currently used by us. It is important to note that the pretrained
NNE's estimates in applications (Section \ref{sec:applications})
are not sensitive to the specification of this distribution.

First, we draw the size of $\boldsymbol{x}$. The size is determined
by $d^{(prod)}$, $d^{(ads)}$, $d^{(cons)}$, $J$, and $n$. These
numbers are picked at random from their respective ranges to which
we would like the pretrained NNE to be applicable. Our current choices
of these ranges are given in Table \ref{tab:data-dim} (in Section
\ref{subsec:search-data}). In particular, the upper bound of $n$
in pretraining is currently set at 2e4. However, this cap does not
restrict the applications of pretrained NNE, because we can always
split a larger dataset and average the estimates across the splits
(also discussed in Section \ref{subsec:search-data}).

Next, we draw the attributes in $\boldsymbol{x}$. In this regard,
we incorporate several features of $\boldsymbol{x}$ commonly seen
in real applications. The first feature is that many attributes show
substantial consumer-level variations. For example, some consumers
see prices systematically higher than other consumers. Therefore,
we draw $\boldsymbol{x}$ with two components: a within-consumer component
plus a cross-consumer component (the latter is also known as consumer
fixed effects). The within-consumer component is ignored for $\boldsymbol{x}_{i}^{(cons)}$,
which by definition varies only cross consumers.

The second feature is that individual attributes often follow non-normal
distributions. To this end, we draw each attribute initially as normal
and then transform it into one of the following non-normal distributions:
(i) a dummy, (ii) a scale from 1 to 5, and (iii) a bell-shaped but
skewed continuous distribution. In principle, more types of distributions
could be included. However, when we do so, the pretrained NNE's estimates
in applications (Section \ref{sec:applications}) remain mostly unchanged.
So these three types of distributions seem sufficient.

The third feature is that attributes are often correlated. To this
end, we let the initial attribute draws follow multivariate normal
distributions with randomly drawn correlation matrices. The correlations
persist through subsequent transformation into non-normal distributions.
For example, if two attributes are positively correlated when first
drawn as normal, they will remain positively correlated when one of
them is transformed into a dummy.

\section{Applications\protect\label{sec:applications}}

We evaluate the pretrained NNE's performance with 12 real datasets.
First, we briefly describe each dataset. Then, we conduct Monte Carlo
studies that estimate the search model on real-data attributes but
simulated consumer choices. Finally, we estimate the search model
on the real data entirely.

Before proceeding, we briefly describe the existing approaches to
estimate search models, for readers unfamiliar with the literature.
Standard SMLE with accept-reject simulations is infeasible for search
models because it requires prohibitive computational costs. A common
workaround in literature is to smooth the simulated likelihood function
(e.g., \citealt{honka2016simultaneous}, \citealt{ursu2018power}).
However, estimates are sensitive to smoothing factors and tuning these
factors requires costly Monte Carlo experiments. More recently, \citet{jiang2021consumer}
and \citet{chung2023simulated} extend the Geweke-Hajivassiliou-Keane
(GHK) simulation algorithm to the likelihood of search models. The
GHK-based SMLE does not require smoothing, but is much more complex
to implement. In this section, we implement both the smoothed SMLE
and the GHK-based SMLE.

\subsection{Datasets\protect\label{subsec:app-data}}

Our 12 datasets come from four different sources (an anonymous online
retailer, Expedia, JD, and Trivago). Table \ref{tab:data-stat} provides
summary statistics. We see that these datasets vary substantially
in the number of search sessions, numbers of attributes, buy rate,
and search rate. So they cover a fairly wide range of real-world scenarios.

For ease of presentation, the estimates reported in this section assume
that each attribute is mean-zero with a standard deviation of 0.2.
The 0.2 standard deviation (instead of 1) makes coefficients ($\boldsymbol{\beta}$,
$\boldsymbol{\eta}$, $\boldsymbol{\alpha}$) larger and thus more
visible in figures. For estimation, all attributes are standardized
to unit variance before being fed into the pretrained NNE (see Section
\ref{subsec:search-data}).

We briefly describe each dataset below.

\begin{table}
\caption{\protect\label{tab:data-stat}Summary Statistics of 12 Datasets}
\smallskip{}

\begin{centering}
\begin{tabular}{lrcccccc}
\hline 
 & $\quad$ & $n$ & $d^{(prod)}$ & $d^{(cons)}$ & $d^{(ads)}$ & buy rate & search rate\tabularnewline
\cline{3-8}
 &  &  &  &  &  &  & \tabularnewline[-3mm]
eCommerce: laptop &  & 10,000 & 6 & 0 & 0 & 5.9\% & 21.5\%\tabularnewline
eCommerce: vacuum &  & 10,000 & 4 & 0 & 0 & 8.5\% & 19.0\%\tabularnewline
eCommerce: washer &  & 10,000 & 5 & 0 & 0 & 7.8\% & 22.1\%\tabularnewline
 &  &  &  &  &  &  & \tabularnewline[-3mm]
Expedia: destination 1 &  & 1,258 & 6 & 2 & 1 & 11.5\% & 8.7\%\tabularnewline
Expedia: destination 2 &  & 897 & 6 & 2 & 1 & 3.8\% & 6.0\%\tabularnewline
 &  &  &  &  &  &  & \tabularnewline[-3mm]
Expedia: Cancun &  & 2,135 & 6 & 0 & 1 & 1.3\% & 30.8\%\tabularnewline
Expedia: Manhattan &  & 1,369 & 6 & 0 & 1 & 3.1\% & 28.7\%\tabularnewline
 &  &  &  &  &  &  & \tabularnewline[-3mm]
JD: desktop channel &  & 5,896 & 4 & 2 & 0 & 16.4\% & 26.5\%\tabularnewline
JD: mobile channel &  & 4,069 & 4 & 2 & 0 & 11.2\% & 12.2\%\tabularnewline
JD: app channel &  & 10,000 & 4 & 2 & 0 & 8.0\% & 12.9\%\tabularnewline
 &  &  &  &  &  &  & \tabularnewline[-3mm]
Trivago: desktop channel &  & 2,082 & 7 & 0 & 1 & 2.8\% & 40.0\%\tabularnewline
Trivago: mobile channel &  & 2,726 & 7 & 0 & 1 & 2.5\% & 29.1\%\tabularnewline
\hline 
\end{tabular}
\par\end{centering}
\smallskip{}

{\footnotesize Notes: $n$ is the number of consumers or search sessions.
$d^{(prod)}$, $d^{(cons)}$, and $d^{(ads)}$ are the numbers of
product attributes, consumer attributes, and advertising attributes,
respectively. Buy rate is the percentage of consumers who made a purchase.
Search rate is the percentage of consumers who made at least one non-free
search.}{\footnotesize\par}
\end{table}

\paragraph*{(1) eCommerce: laptop}

Our first data source consists of consumer searches and purchases
in an anonymous large online store, provided by the marketing platform
REES46. The data are available on Kaggle.com and include many product
categories.\footnote{https://www.kaggle.com/datasets/mkechinov/ecommerce-behavior-data-from-multi-category-store}
We focus on laptops in this dataset. The product attributes $\boldsymbol{x}_{ij}^{(prod)}$
include price and five brand dummies: Asus, HP, Lenovo, MSI, and Xiaomi
(the baseline brand is Acer). There are no consumer or advertising
attributes. There are a very large number of search sessions. To ease
the computational burden of SMLE, we use a subsample of 10,000 search
sessions.

\paragraph*{(2) eCommerce: vacuum}

This dataset includes the vacuum category in the anonymous online
store. The product attributes $\boldsymbol{x}_{ij}^{(prod)}$ include
price and three brand dummies: Philips, Samsung, and Vitek (the baseline
brand is Bosch). We again take a subsample of 10,000 search sessions.

\paragraph*{(3) eCommerce: washer}

This dataset includes the washer category in the anonymous online
store. The product attributes $\boldsymbol{x}_{ij}^{(prod)}$ include
price and four brand dummies: Beko, Candy, Indesit, Midea (the baseline
brand is Atlant). We again take a subsample of 10,000 search sessions.

\paragraph{(4) Expedia: destination 1}

The International Conference on Data Mining (ICDM 2013) organized
a contest to predict consumer searches and bookings for hotels on
Expedia. The data for the contest are publicly available on Kaggle.com.\footnote{https://www.kaggle.com/competitions/expedia-personalized-sort/overview}
Several papers have used the data to study consumer search behaviors
(\citealt{ursu2018power}, \citealt{greminger2022heterogeneous},
\citealt{compiani2023online}). The data include multiple travel destinations.
``Destination 1'' refers to the largest destination. The product
attributes $\boldsymbol{x}_{ij}^{(prod)}$ include star rating, review
score, a chain indicator, location score, price, and promotion. The
consumer attributes $\boldsymbol{x}_{i}^{(cons)}$ include a weekend
dummy and a working-hours dummy. The advertising attribute $\boldsymbol{x}_{ij}^{(ads)}$
includes the ranking of the hotel.

\paragraph{(5) Expedia: destination 2}

This dataset consists of the search sessions for the second-largest
destination in the Expedia data from ICDM 2013. There are only $n=897$
search sessions, slightly below the lower bound of $n$ set for the
pretrained NNE (see Table \ref{tab:data-dim}). Nonetheless, we choose
to include the dataset, because it allows us to test the pretrained
NNE slightly outside its intended range of applications.

\paragraph{(6) Expedia: Cancun}

The Analytics at Wharton Initiative also provides data on hotel searches
and bookings on Expedia.\footnote{https://analytics.wharton.upenn.edu/open-research-opportunities/predicting-optimizing-consumer-response-to-product-search-results/}
The data are not openly available but can be accessed after a request
with the Initiative. The data include four travel destinations (Cancun,
Manhattan, Paris and Budapest). Cancun is the largest destination.

\paragraph{(7) Expedia: Manhattan}

This dataset consists of the search sessions for the second-largest
destination (Manhattan) in the Expedia data from Analytics at Wharton.

\paragraph{(8) JD: desktop channel}

The 2020 MSOM Data Driven Research Challenge provided data on consumer
experiences on JD.com. The data were made available to INFORMS members
in 2019.\footnote{https://connect.informs.org/msom/events/datadriven2020}
JD.com is China\textquoteright s second largest business-to-consumer
online retail platform. The data cover many aspects of consumer experiences
on the platform, from website browsing to order fulfillment (see \citealt{shen2020jd}
for details). We focus on consumer search and purchase activities.
The consumer attributes $\boldsymbol{x}_{i}^{(cons)}$ include a weekend
dummy and a working-hours dummy. The product attributes $\boldsymbol{x}_{ij}^{(prod)}$
include price, a dummy for first-party sellers, and two unnamed attributes.
There are no advertising attributes.

The JD.com data include three channels: desktop, mobile, and app.
This dataset focuses on the desktop channel.

\paragraph{(9) JD: mobile channel }

This dataset includes the mobile channel from the JD.com data.

\paragraph{(10) JD: app channel}

This dataset includes the app channel from the JD.com data. There
are a large number of search sessions. To ease the computational burden
of SMLE, we use a subsample of 10,000 search sessions.

\paragraph{(11) Trivago: desktop channel}

The ACM RecSys Challenge 2019 provided consumer browsing data on Trivago.com.\footnote{https://recsys.acm.org/recsys19/challenge/}
The data are not standard for search models because purchases are
not very well defined in this context, as we will explain below. Nonetheless,
we choose to include the data because doing so allows us to test the
pretrained NNE slightly outside its intended context of applications.

Trivago is a hotel meta-search engine that aggregates information
from hotels and travel websites (similar to how Google Flights aggregates
flight information). On Trivago.com, consumers receive a list of hotels.
They can click on any hotel to see more information. They can also
click a \textquotedblleft View Deal\textquotedblright{} button that
directs them to an external website for possible booking. But what
happens at external websites is not observed in the data. We treat
a click for more information as a search and a \textquotedblleft View
Deal\textquotedblright{} click as a purchase. Sometimes a consumer
clicks ``View Deal'' for multiple hotels, in which case we randomly
pick one as the purchase. Cases where consumers click ``View Deal''
without first clicking for more information are dropped (about 5\%).

The Trivago data include search sessions from both the desktop channel
and mobile channel. This dataset focuses on the desktop channel.

\paragraph{(12) Trivago: mobile channel}

This dataset consists of the search sessions from the mobile channel
in the Trivago data.

\subsection{Monte Carlo studies\protect\label{subsec:applications-MC}}

\begin{table}
\setlength\tabcolsep{3pt}

\caption{\protect\label{tab:mc-par}Estimates in Monte Carlo Study with eCommerce:
Laptop}
\medskip{}

\begin{centering}
\begin{tabular}{rrrrlrlrlrlrl}
\hline 
 & Truth & $\quad$ & \multicolumn{2}{>{\centering}p{7em}}{Pretrained NNE} & \multicolumn{2}{>{\centering}p{7em}}{SMLE\\
 (GHK)\\
$R=50$} & \multicolumn{2}{>{\centering}p{7em}}{SMLE\\
 (smoothed)\\
$R=50$ $\boldsymbol{\lambda}=(4,1)$} & \multicolumn{2}{>{\centering}p{7em}}{SMLE\\
 (smoothed)\\
$R=50$ $\boldsymbol{\lambda}=(2,1)$} & \multicolumn{2}{>{\centering}p{7em}}{SMLE\\
 (smoothed)\\
$R=50$ $\boldsymbol{\lambda}=(4,2)$}\tabularnewline
\hline 
 &  &  &  &  &  &  &  &  &  &  &  & \tabularnewline[-3mm]
$\alpha_{0}$ & -5.0 &  & -5.009 & {\scriptsize (0.006)} & -5.779 & {\scriptsize (0.010)} & -4.696 & {\scriptsize (0.023)} & -3.347 & {\scriptsize (0.011)} & -2.767 & {\scriptsize (0.011)}\tabularnewline
$\eta_{0}$ & 4.8 &  & 4.797 & {\scriptsize (0.002)} & 4.732 & {\scriptsize (0.004)} & 4.851 & {\scriptsize (0.009)} & 4.981 & {\scriptsize (0.005)} & 3.999 & {\scriptsize (0.006)}\tabularnewline
$\beta_{1}$ & -0.6 &  & -0.603 & {\scriptsize (0.002)} & -0.579 & {\scriptsize (0.003)} & -0.375 & {\scriptsize (0.005)} & -0.457 & {\scriptsize (0.005)} & -0.366 & {\scriptsize (0.005)}\tabularnewline
$\beta_{2}$ & -0.5 &  & -0.488 & {\scriptsize (0.002)} & -0.485 & {\scriptsize (0.003)} & -0.300 & {\scriptsize (0.005)} & -0.370 & {\scriptsize (0.005)} & -0.295 & {\scriptsize (0.005)}\tabularnewline
$\beta_{3}$ & -0.3 &  & -0.292 & {\scriptsize (0.002)} & -0.290 & {\scriptsize (0.003)} & -0.188 & {\scriptsize (0.005)} & -0.228 & {\scriptsize (0.006)} & -0.183 & {\scriptsize (0.005)}\tabularnewline
$\beta_{4}$ & -1.0 &  & -1.010 & {\scriptsize (0.003)} & -0.968 & {\scriptsize (0.003)} & -0.627 & {\scriptsize (0.005)} & -0.771 & {\scriptsize (0.005)} & -0.616 & {\scriptsize (0.005)}\tabularnewline
$\beta_{5}$ & -0.4 &  & -0.399 & {\scriptsize (0.003)} & -0.390 & {\scriptsize (0.003)} & -0.255 & {\scriptsize (0.005)} & -0.315 & {\scriptsize (0.006)} & -0.250 & {\scriptsize (0.005)}\tabularnewline
$\beta_{6}$ & -0.5 &  & -0.489 & {\scriptsize (0.003)} & -0.479 & {\scriptsize (0.003)} & -0.316 & {\scriptsize (0.006)} & -0.392 & {\scriptsize (0.006)} & -0.312 & {\scriptsize (0.005)}\tabularnewline
\hline 
\end{tabular}\medskip{}
\par\end{centering}
{\footnotesize Notes: $R$ is the number of simulations per consumer
in SMLE. $\boldsymbol{\lambda}$ is the smoothing factors in the smoothed
SMLE. Results are averaged across 100 Monte Carlo repetitions. Numbers
in parentheses are the standard errors for the averages.}{\footnotesize\par}
\end{table}

Our first set of tests are Monte Carlo studies. Specifically, for
each of the 12 real datasets, we: (i) keep the product, consumer,
and advertising attributes, but (ii) replace the search and purchase
choices with the choices simulated by the search model under a ``true''
$\boldsymbol{\theta}$. Then, we try to recover $\boldsymbol{\theta}$
from this dataset. Knowing the ground truth of $\boldsymbol{\theta}$
allows us to measure estimation accuracy. We set the true $\boldsymbol{\theta}$
close to its estimate from Section \ref{subsec:applications-real}.

\begin{figure}
\caption{\protect\label{fig:MC-smoothing} Monte Carlo Study with eCommerce:
Laptop}

\vspace{2em}

\begin{centering}
\noindent\begin{minipage}[t]{1\columnwidth}%
\begin{center}
\includegraphics[scale=0.85]{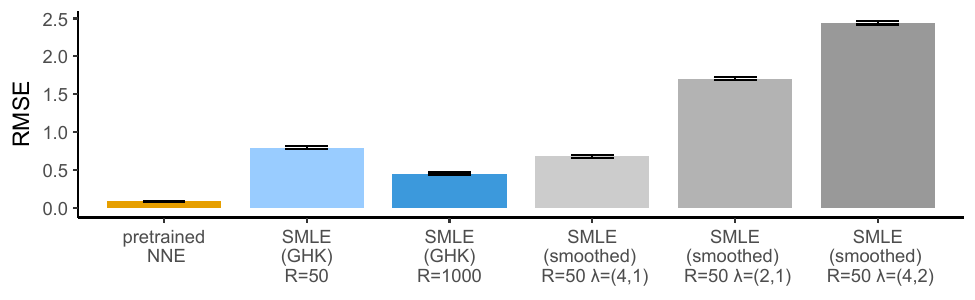}
\par\end{center}%
\end{minipage}\vspace{2em}
\par\end{centering}
\begin{centering}
\noindent\begin{minipage}[t]{1\columnwidth}%
\begin{center}
\includegraphics[scale=0.85]{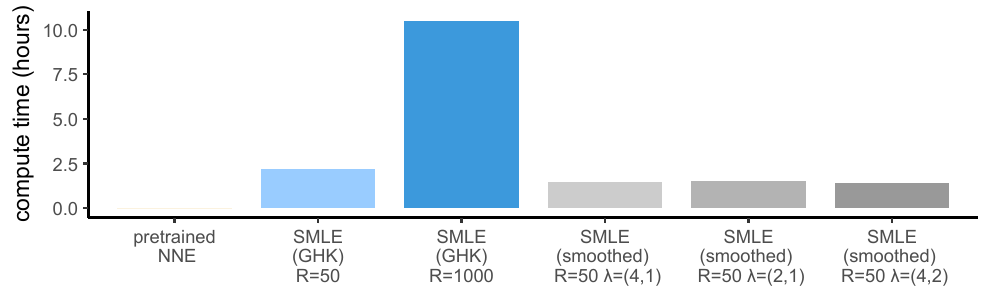}
\par\end{center}%
\end{minipage}
\par\end{centering}
{\footnotesize Notes: Results are based on 100 Monte Carlo repetitions. }{\footnotesize\par}
\end{figure}

Below, we first closely examine the results from the eCommerce: laptop
dataset. In Table \ref{tab:mc-par}, Column 1 shows the true value
of $\boldsymbol{\theta}$, Column 2 shows the pretrained NNE's estimate,
and Columns 3 - 6 show the SMLE's estimates. The $R$ denotes the
number of simulations per consumer in SMLE and $R=50$ is typical
in literature. For the smoothed SMLE, $\boldsymbol{\lambda}$ collects
the smoothing factors for search choices and for purchase choices.\footnote{Some applications use the same smoothing factor for search and purchase
choices. But doing so here worsens RMSE substantially. Whether multiple
smoothing factors are necessary is generally application-dependent.} The table shows three settings of $\boldsymbol{\lambda}$ to illustrate
the sensitivity of SMLE's estimate to $\boldsymbol{\lambda}$. The
optimal $\boldsymbol{\lambda}$ is generally application-dependent.
We actually have conducted an expensive grid search of $\boldsymbol{\lambda}$,
and $\boldsymbol{\lambda}=(4,1)$ gives the smallest root mean square
error (RMSE) in this application.

The upper plot of Figure \ref{fig:MC-smoothing} reports the RMSEs
of the different estimators. The pretrained NNE has the lowest RMSE,
outperforming even the GHK-based SMLE that uses a very large $R=1000$.
Although we generally expect MLE to be asymptotically efficient, the
performance of SMLE is affected by simulation errors. In search models,
simulating the likelihood is particularly difficult because of the
large number of possible search and purchase combinations. Our result
here suggests that simulating this likelihood remains challenging
despite workarounds such as smoothing and GHK. The result also shows
that pretrained NNE can achieve high estimation accuracy despite being
off-the-shelf.

The lower plot of Figure \ref{fig:MC-smoothing} reports the computation
time needed for a single estimation run. All the times are measured
on the same hardware. Recall that this dataset is of moderate size
(a subsample from a single product category; see Section \ref{subsec:app-data}).\footnote{There are 624 categories in the original data. The laptop category
has 41,918 search sessions. Our dataset is a subsample of 10,000 search
sessions.} Even so, the SMLE requires between 1.5 and 10 hours. The pretrained
NNE takes about 1 second (so the orange bar is barely visible). This
represents a time saving of 3 to 4 orders of magnitude. In addition,
we note that implementing SMLE requires additional costs not reflected
in Figure \ref{fig:MC-smoothing}, such as tuning smoothing factors,
tuning optimization routines, trying starting points. Each of these
tasks requires multiple estimation runs. Moreover, the GHK takes considerable
effort to code. While pretraining an NNE also has its own implementation
costs, they arise only once during pretraining and are not borne by
the subsequent users of the pretrained NNE.

Next, we expand the examination to all 12 datasets. Table \ref{tab:MC-all}
reports the results. The table omits the smoothed SMLE because, as
we have seen, its estimates are sensitive to the smoothing factors
$\boldsymbol{\lambda}$ (and the optimal $\boldsymbol{\lambda}$ is
application-dependent). The main finding with Table \ref{tab:MC-all}
is qualitatively the same as what we have seen with the eCommerce:
laptop dataset. The pretrained NNE achieves either comparable or better
estimation accuracy at drastically lower computational costs.

Finally, it is worth noting that the 12 datasets here are not large
(see Table \ref{tab:data-stat}). In fact, recall that each dataset
focuses on only a subset (a single product category, travel destination,
or sales channel) of its source, and four of the 12 datasets are subsampled
to reduce the computation burden. Because computation time scales
with data size, the absolute time savings on large datasets will be
much greater than those indicated by Table \ref{tab:MC-all}.

\begin{table}
\setlength\tabcolsep{4pt}\caption{\protect\label{tab:MC-all}Monte Carlo Studies Across 12 Datasets}
\medskip{}

\begin{centering}
\begin{tabular}{llrlrrlrrl}
\hline 
 & $\quad$ & \multicolumn{2}{>{\raggedleft}p{6em}}{Pretrained NNE} & $\quad$ & \multicolumn{2}{>{\raggedleft}p{6em}}{SMLE (GHK) $R=50$} & $\quad$ & \multicolumn{2}{>{\raggedleft}p{6em}}{SMLE (GHK) $R=1000$}\tabularnewline
\hline 
 & \multicolumn{9}{c}{RMSE}\tabularnewline
\cline{2-10}
 &  &  &  &  &  &  &  &  & \tabularnewline[-3mm]
eCommerce: laptop &  & 0.094 & {\scriptsize (0.003)} &  & 0.795 & {\scriptsize (0.010)} &  & 0.453 & {\scriptsize (0.009)}\tabularnewline
eCommerce: vacuum &  & 0.095 & {\scriptsize (0.003)} &  & 0.352 & {\scriptsize (0.005)} &  & 0.176 & {\scriptsize (0.005)}\tabularnewline
eCommerce: washer &  & 0.179 & {\scriptsize (0.009)} &  & 0.614 & {\scriptsize (0.007)} &  & 0.361 & {\scriptsize (0.008)}\tabularnewline
 &  &  &  &  &  &  &  &  & \tabularnewline[-3mm]
Expedia: destination 1 &  & 0.451 & {\scriptsize (0.012)} &  & 0.591 & {\scriptsize (0.022)} &  & 0.501 & {\scriptsize (0.015)}\tabularnewline
Expedia: destination 2 &  & 0.748 & {\scriptsize (0.023)} &  & 0.964 & {\scriptsize (0.035)} &  & 0.833 & {\scriptsize (0.028)}\tabularnewline
 &  &  &  &  &  &  &  &  & \tabularnewline[-3mm]
Expedia: Cancun &  & 0.485 & {\scriptsize (0.021)} &  & 1.814 & {\scriptsize (0.055)} &  & 1.083 & {\scriptsize (0.045)}\tabularnewline
Expedia: Manhattan &  & 0.500 & {\scriptsize (0.018)} &  & 1.449 & {\scriptsize (0.054)} &  & 0.855 & {\scriptsize (0.040)}\tabularnewline
 &  &  &  &  &  &  &  &  & \tabularnewline[-3mm]
JD: desktop channel &  & 0.175 & {\scriptsize (0.005)} &  & 0.392 & {\scriptsize (0.008)} &  & 0.256 & {\scriptsize (0.008)}\tabularnewline
JD: mobile channel &  & 0.234 & {\scriptsize (0.007)} &  & 0.375 & {\scriptsize (0.011)} &  & 0.273 & {\scriptsize (0.009)}\tabularnewline
JD: app channel &  & 0.161 & {\scriptsize (0.006)} &  & 0.335 & {\scriptsize (0.006)} &  & 0.200 & {\scriptsize (0.008)}\tabularnewline
 &  &  &  &  &  &  &  &  & \tabularnewline[-3mm]
Trivago: desktop channel &  & 0.423 & {\scriptsize (0.016)} &  & 2.462 & {\scriptsize (0.050)} &  & 1.521 & {\scriptsize (0.044)}\tabularnewline
Trivago: mobile channel &  & 0.422 & {\scriptsize (0.021)} &  & 1.067 & {\scriptsize (0.037)} &  & 0.660 & {\scriptsize (0.030)}\tabularnewline
\hline 
 & \multicolumn{9}{c}{Compute time (mins)}\tabularnewline
\cline{2-10}
 &  &  &  &  &  &  &  &  & \tabularnewline[-3mm]
eCommerce: laptop &  & 0.02 &  &  & 132.72 &  &  & 630.06 & \tabularnewline
eCommerce: vacuum &  & 0.02 &  &  & 59.42 &  &  & 371.52 & \tabularnewline
eCommerce: washer &  & 0.02 &  &  & 126.48 &  &  & 552.85 & \tabularnewline
 &  &  &  &  &  &  &  &  & \tabularnewline[-3mm]
Expedia: destination 1 &  & 0.01 &  &  & 20.35 &  &  & 126.20 & \tabularnewline
Expedia: destination 2 &  & 0.01 &  &  & 17.53 &  &  & 94.03 & \tabularnewline
 &  &  &  &  &  &  &  &  & \tabularnewline[-3mm]
Expedia: Cancun &  & 0.02 &  &  & 27.13 &  &  & 151.97 & \tabularnewline
Expedia: Manhattan &  & 0.02 &  &  & 19.01 &  &  & 101.17 & \tabularnewline
 &  &  &  &  &  &  &  &  & \tabularnewline[-3mm]
JD: desktop channel &  & 0.02 &  &  & 50.43 &  &  & 306.65 & \tabularnewline
JD: mobile channel &  & 0.01 &  &  & 33.53 &  &  & 213.45 & \tabularnewline
JD: app channel &  & 0.02 &  &  & 85.27 &  &  & 495.20 & \tabularnewline
 &  &  &  &  &  &  &  &  & \tabularnewline[-3mm]
Trivago: desktop channel &  & 0.02 &  &  & 33.98 &  &  & 212.17 & \tabularnewline
Trivago: mobile channel &  & 0.01 &  &  & 46.50 &  &  & 281.87 & \tabularnewline
\hline 
\end{tabular} 
\par\end{centering}
\medskip{}

{\footnotesize Notes: $R$ is the number of simulations per consumer
in SMLE. $R=50$ is typical in literature. Results are averaged across
100 Monte Carlo repetitions. Numbers in parentheses are the standard
errors for the averages.}{\footnotesize\par}
\end{table}

\subsection{Real-data estimates \protect\label{subsec:applications-real}}

\begin{figure}
\caption{\protect\label{fig:real-estimates-all}Real-Data Estimates Across
12 Datasets}
\vspace{2em}

\begin{centering}
\begin{minipage}[t]{0.33\textwidth}%
\begin{center}
{\footnotesize eCommerce: laptop\vspace{-2em}}{\footnotesize\par}
\par\end{center}
\begin{center}
\includegraphics[scale=0.85]{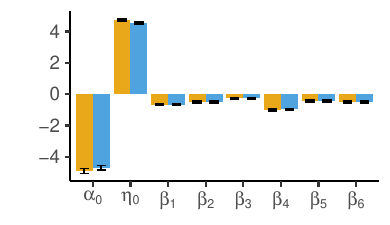}
\par\end{center}%
\end{minipage}%
\begin{minipage}[t]{0.33\textwidth}%
\begin{center}
{\footnotesize eCommerce: vacuum\vspace{-2em}}{\footnotesize\par}
\par\end{center}
\begin{center}
\includegraphics[scale=0.85]{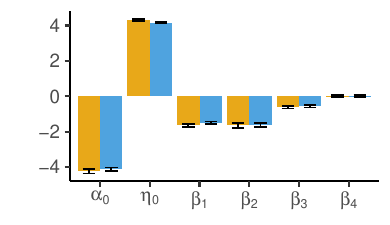}
\par\end{center}%
\end{minipage}%
\begin{minipage}[t]{0.33\textwidth}%
\begin{center}
{\footnotesize eCommerce: washer\vspace{-2em}}{\footnotesize\par}
\par\end{center}
\begin{center}
\includegraphics[scale=0.85]{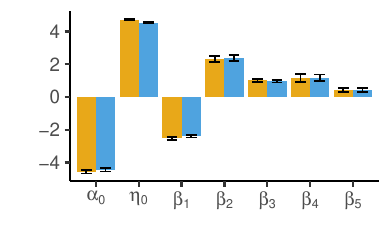}
\par\end{center}%
\end{minipage}
\par\end{centering}
\vspace{1.75em}

\begin{centering}
\begin{minipage}[t]{0.33\textwidth}%
\begin{center}
{\footnotesize Expedia: destination 1\vspace{-2em}}{\footnotesize\par}
\par\end{center}
\begin{center}
\includegraphics[scale=0.85]{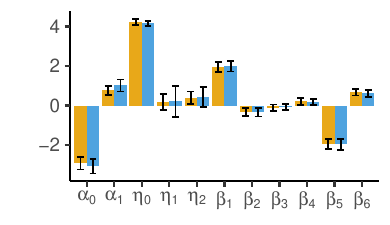}
\par\end{center}%
\end{minipage}%
\begin{minipage}[t]{0.33\textwidth}%
\begin{center}
{\footnotesize Expedia: destination 2\vspace{-2em}}{\footnotesize\par}
\par\end{center}
\begin{center}
\includegraphics[scale=0.85]{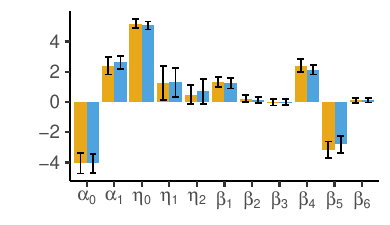}
\par\end{center}%
\end{minipage}%
\begin{minipage}[t]{0.33\textwidth}%
\begin{center}
{\footnotesize Expedia: Cancun\vspace{-2em}}{\footnotesize\par}
\par\end{center}
\begin{center}
\includegraphics[scale=0.85]{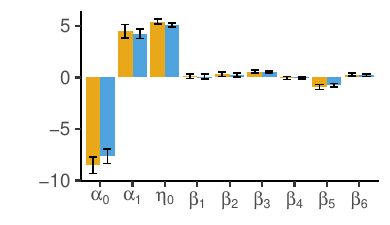}
\par\end{center}%
\end{minipage}\vspace{1.75em}
\par\end{centering}
\begin{centering}
\begin{minipage}[t]{0.33\textwidth}%
\begin{center}
{\footnotesize Expedia: Manhattan\vspace{-2em}}{\footnotesize\par}
\par\end{center}
\begin{center}
\includegraphics[scale=0.85]{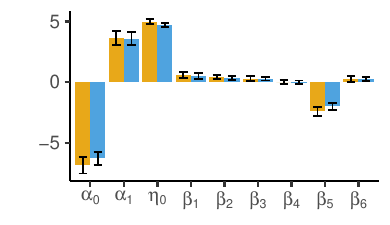}
\par\end{center}%
\end{minipage}%
\begin{minipage}[t]{0.33\textwidth}%
\begin{center}
{\footnotesize JD: desktop\vspace{-2em}}{\footnotesize\par}
\par\end{center}
\begin{center}
\includegraphics[scale=0.85]{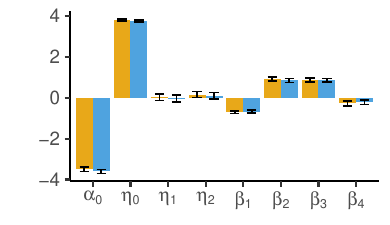}
\par\end{center}%
\end{minipage}%
\begin{minipage}[t]{0.33\textwidth}%
\begin{center}
{\footnotesize JD: mobile\vspace{-2em}}{\footnotesize\par}
\par\end{center}
\begin{center}
\includegraphics[scale=0.85]{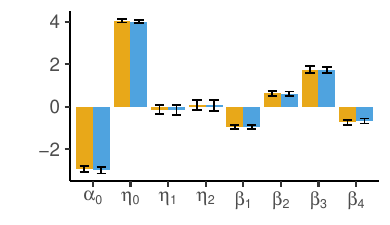}
\par\end{center}%
\end{minipage}
\par\end{centering}
\begin{centering}
\vspace{1.75em}
\begin{minipage}[t]{0.33\textwidth}%
\begin{center}
{\footnotesize JD: app\vspace{-2em}}{\footnotesize\par}
\par\end{center}
\begin{center}
\includegraphics[scale=0.85]{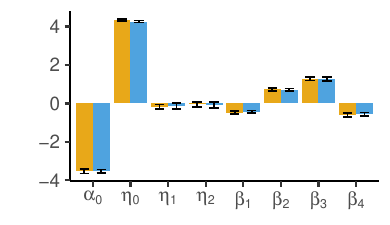}
\par\end{center}%
\end{minipage}%
\begin{minipage}[t]{0.33\textwidth}%
\begin{center}
{\footnotesize Trivago: desktop\vspace{-2em}}{\footnotesize\par}
\par\end{center}
\begin{center}
\includegraphics[scale=0.85]{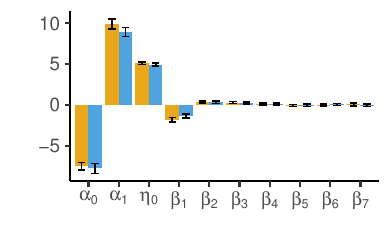}
\par\end{center}%
\end{minipage}%
\begin{minipage}[t]{0.33\textwidth}%
\begin{center}
{\footnotesize Trivago: mobile\vspace{-2em}}{\footnotesize\par}
\par\end{center}
\begin{center}
\includegraphics[scale=0.85]{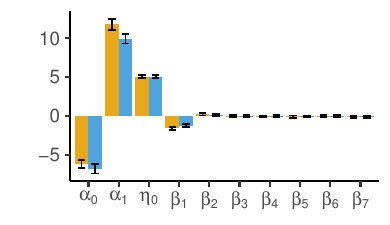}
\par\end{center}%
\end{minipage}\vspace{1.5em}
\par\end{centering}
\centering{}%
\begin{minipage}[t]{0.33\textwidth}%
\begin{center}
\includegraphics[scale=0.85]{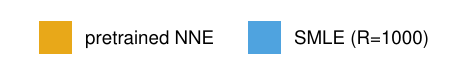}
\par\end{center}%
\end{minipage}
\end{figure}

Our second set of tests estimates the search model entirely from real
data -- that is, both real attributes and real consumer choices.
We no longer have a ground truth for $\boldsymbol{\theta}$ and therefore
cannot measure estimation accuracy directly. Instead, we assess whether
the pretrained NNE's estimates are sensible, by comparing them with
the estimates from the GHK-based SMLE that uses the very large $R=1000$.

Figure \ref{fig:real-estimates-all} shows the estimates across the
12 datasets. Each plot corresponds to one dataset. The horizontal
axis lists the individual parameters in $\boldsymbol{\theta}$. The
orange bars show the pretrained NNE's estimates, while the blue bars
show the SMLE's estimates. A 95\% confidence interval is shown on
each bar. Overall, we see that the estimates from the pretrained NNE
and from the SMLE are closely matched.

We omit reporting computation times in Figure \ref{fig:real-estimates-all},
because they are quantitatively similar to those reported in Table
\ref{tab:MC-all} for Monte Carlo studies. Across all the datasets,
the pretrained NNE uses only a fraction of the computation time required
by SMLE.

\section{Conclusion\protect\label{sec:conclusion}}

This paper explores the feasibility of pretraining an off-the-shelf
machine learning model to estimate a given structural econometric
model. We do not use the machine learning model to capture relationships
in data. Instead, these relationships are prescribed by the structural
model, and we use the machine learning model to recover this structural
model's parameter. Pretrained NNE enables users to estimate a structural
model as easily as reduced-form regressions, while providing equal
or better accuracy than existing estimation methods (at least in the
context of the search model).

Structural estimation is very valuable in empirical research ---
it offers insights into micro-founded consumer preferences and enables
counterfactual analyses. However, its wider use has been limited by
the large amount of effort and time required to estimate structural
models. Pretrained estimators remove these barriers, making structural
models more accessible to both researchers and practitioners.

Further, the ease of computation offered by pretrained NNE may facilitate
new applications. One notable area is to integrate structural models
with real-time algorithms, such as dynamic pricing, ranking algorithms,
recommender systems, and bandit experiments (e.g., \citealt{agrawal2019mnl},
\citealt{jain2024effective}). Because structural models provide estimates
of latent consumer preferences, these estimates can be fed into real-time
algorithms to improve their decisions, e.g., what price to set or
what items to recommend. In addition, counterfactual analyses can
be used to evaluate and provide feedback on the algorithms. However,
in such applications the structural models must be estimated quickly
in real time. Pretrained NNE offers a way to achieve such fast estimation.

The intention of this paper is to explore the feasibility of pretrained
estimators, rather than to develop a comprehensive estimator for search
models. This leaves areas where the pretrained NNE can be further
improved. We note two potential areas. The first area is to increase
the richness of the structural model. The search model considered
in this paper is a standard yet simple one. Future work may allow
for attributes that simultaneously affect utility and search costs,
or allow for richer unobserved heterogeneity in utility and search
costs. These extensions will require us to find additional reduced-form
patterns to help identify the more complex structural model.

A second area for improvement is to address potential endogeneity.
For example, if researchers have an instrumental variable (IV) for
price, how do we build a pretrained estimator that can leverage the
instrument? At a high level, addressing this problem requires two
tasks. The first task is to generate training datasets that include
instruments. The second task is to pass relevant information regarding
the instrument to the neural net. These tasks are not straightforward.
However, incorporating IV in NNE will be a significant and valuable
extension, because we frequently need to address endogeneity in structural
estimation applications.

\appendix
\newpage
\renewcommand*{\bibfont}{\small}

\bibliographystyle{econ}
\bibliography{reference}

\appendix
\newpage

\section*{Appendix }

\section{Data patterns\protect\label{sec:app-moments}}

Section \ref{subsec:pnne-overview} summarized the data patterns $\boldsymbol{m}$
that we feed into the neural net. Below, we provide the exact specification
of $\boldsymbol{m}$.

We first recall some notation that will be needed. Vectors $\boldsymbol{x}_{ij}^{(prod)}$,
$\boldsymbol{x}_{ij}^{(ads)}$, and $\boldsymbol{x}_{i}^{(cons)}$
collect the product, advertising, and consumer attributes, respectively.
It is assumed that all attributes are standardized. As before, let
$d^{(prod)}$, $d^{(ads)}$, and $d^{(cons)}$ denote the numbers
of attributes. These numbers are dataset-dependent. Let $d_{max}^{(prod)}$,
$d_{max}^{(ads)}$, and $d_{max}^{(cons)}$ denote the maximum values
that we currently set for the numbers of attributes (given in Table
\ref{tab:data-dim}). These maximum values are not dataset-dependent.

We now define several consumer-level variables that will be used.
Recall that $J$ is the number of products per consumer and $n$ is
the number of consumers.
\begin{itemize}
\item Let $\bar{\boldsymbol{x}}_{i}^{(prod)}=\frac{1}{J}\sum_{j=1}^{J}\boldsymbol{x}_{ij}^{(prod)}$
denote the average product attributes for consumer $i$.
\item Let $\bar{\boldsymbol{x}}_{i}^{(ads)}=\frac{1}{J}\sum_{j=1}^{J}\boldsymbol{x}_{ij}^{(ads)}$
denote the average advertising attributes for consumer $i$.
\item Let $\tilde{y}_{i}^{(search)}=\sum_{j=1}^{J}y_{ij}^{(search)}$ denote
the number of searches made by consumer $i$.
\item Let $\tilde{y}_{i}^{(nf)}=\mathbb{I}\big\{\tilde{y}_{i}^{(search)}>1\big\}$
indicate whether consumer $i$ did any non-free search.
\item Let $\tilde{y}_{i}^{(buy)}=\sum_{j=1}^{J}y_{ij}^{(buy)}$, which is
a dummy indicating whether consumer $i$ made a purchase.
\end{itemize}
The elements of $\boldsymbol{m}$ for any given dataset ${\cal D}$
are:
\begin{enumerate}
\item The buy rate, search rate, and average number of searches per consumer.
\item The values of $J$ and $n$.
\item A dummy vector indicating the numbers of product, advertising, and
consumer attributes, shown as follows. Note that the vector has a
fixed length of $d_{max}\equiv d_{max}^{(prod)}+d_{max}^{(ads)}+d_{max}^{(cons)}$.
\begin{equation}
(\quad\underbrace{1,...,1,}_{d^{(prod)}}\quad\underbrace{0,...,0,}_{d_{max}^{(prod)}-d^{(prod)}}\quad\underbrace{1,...,1,}_{d^{(ads)}}\quad\underbrace{0,...,0,}_{d_{max}^{(ads)}-d^{(ads)}}\quad\underbrace{1,...,1,}_{d^{(cons)}}\quad\underbrace{0,...,0}_{d_{max}^{(cons)}-d^{(cons)}}).\label{eq:dummy_vector}
\end{equation}
\item The coefficients for $\{1,\;\boldsymbol{x}_{ij}^{(prod)},\boldsymbol{x}_{ij}^{(ads)},\boldsymbol{x}_{i}^{(cons)}\}$
in a logit regression of $y_{ij}^{(search)}$ on $\{1,\;\boldsymbol{x}_{ij}^{(prod)},\allowbreak\boldsymbol{x}_{ij}^{(ads)},\boldsymbol{x}_{i}^{(cons)},\bar{\boldsymbol{x}}_{i}^{(prod)},\bar{\boldsymbol{x}}_{i}^{(ads)}\}$.
These coefficients are padded with zeros and collected in a vector
of fixed length $d_{max}+1$. The paddings enter this vector in the
same fashion as in expression (\ref{eq:dummy_vector}). This style
of collecting regression coefficients applies to all the regressions
below.
\item The coefficients in a multinomial logit regression of $y_{ij}^{(buy)}$
on $\{1,\;\boldsymbol{x}_{ij}^{(prod)},\boldsymbol{x}_{ij}^{(ads)},\boldsymbol{x}_{i}^{(cons)}\}$,
conditional on $y_{ij}^{(search)}=1$. Intuitively, this regression
describes how each consumer chooses among searched products.
\item The coefficients for $\{1,\;\boldsymbol{x}_{i}^{(cons)}\}$ in a linear
regression of $\log(\tilde{y}_{i}^{(search)})$ on $\{1,\;\boldsymbol{x}_{i}^{(cons)},\bar{\boldsymbol{x}}_{i}^{(ads)},\bar{\boldsymbol{x}}_{i}^{(prod)}\}$.
\item The coefficients for $\{1,\;\boldsymbol{x}_{i}^{(cons)}\}$ in a logit
regression of $\tilde{y}_{i}^{(nf)}$ on $\{1,\;\boldsymbol{x}_{i}^{(cons)},\bar{\boldsymbol{x}}_{i}^{(ads)},\bar{\boldsymbol{x}}_{i}^{(prod)}\}$.
\item The coefficients for $\{1,\boldsymbol{x}_{i}^{(cons)}\}$ in a logit
regression of $\tilde{y}_{i}^{(buy)}$ on $\{1,\;\boldsymbol{x}_{i}^{(cons)},\bar{\boldsymbol{x}}_{i}^{(ads)},\bar{\boldsymbol{x}}_{i}^{(prod)}\}$.
\item The means and standard deviations of the variables in all the regressions
above, except for those with fixed values (e.g., the standard deviations
of $\boldsymbol{x}_{ij}^{(prod)}$ always equal to 1).
\end{enumerate}
\vspace{0.5em}

All the regressions above are estimated using ridge regression that
imposes a small penalty on the coefficients. Using ridge regression
leads to a noticeable decrease in the validation loss of the pretrained
NNE. The validation loss is further decreased when we estimate each
regression with two different penalty levels, and include the coefficients
from both penalty levels in $\boldsymbol{m}$. We currently use the
penalty levels of 0.001 and 1e-6. Using three or more penalty levels
yields little additional decrease in the validation loss.

\section{Detecting ill-suited applications \protect\label{sec:app-trim}}

As discussed, we say a dataset is ill-suited if the associated data
patterns $\boldsymbol{m}$ lies outside the range of data patterns
seen in pretraining. In this case, the neural net has to extrapolate
and its output may not be reliable. Because a pretrained NNE can be
applied at low cost to many datasets, it is helpful to have a way
to detect whether a particular dataset is ill-suited and to inform
users when this is the case.

Specifically, we want to detect whether the $\boldsymbol{m}$ from
a given real dataset is atypical (an outlier) relative to $\{\boldsymbol{m}^{(\ell)}\}_{\ell=1}^{L}$
from the training examples. We note that $\boldsymbol{m}$ has over
a hundred dimensions, and the detection cannot be reduced to checking
each dimension separately. Even if every individual element of $\boldsymbol{m}$
appears typical on its own, the vector $\boldsymbol{m}$ can still
be atypical if it contradicts some relationship between elements that
holds in the training examples.

We propose a practical method of detection by making use of a tree-based
model (boosted tree or random forest). In general, when a tree-based
model and a neural net are trained on the same training examples,
their predictions should not differ substantially within the region
covered by training examples. However, a tree-based model and a neural
net tend to extrapolate very differently. So, let $\boldsymbol{f}^{NET}$
denote the neural net and $\boldsymbol{f}^{TREE}$ denote a tree-based
model, both trained with the same training examples. We know $\boldsymbol{m}$
is atypical if $\boldsymbol{f}^{NET}(\boldsymbol{m})$ and $\boldsymbol{f}^{TREE}(\boldsymbol{m})$
differ substantially. In this case, we display a warning to users.

It is worth pointing out that $\boldsymbol{f}^{NET}$ has a lower
validation loss than $\boldsymbol{f}^{TREE}$. In other words, within
the region covered by training examples, the neural net outperforms
the tree-based model. Therefore, in any application that is not deemed
ill-suited, we use the output from the neural net instead of the tree-based
model.

\section{Proof}

Proposition \ref{prop:convergence} is proved by following \citet{wei2024estimating}'s
proof (which builds on \citealt{chen2007sieve} and \citealt{white1989learning}).
The only difference is that \citet{wei2024estimating} work with a
distribution of $(\boldsymbol{\theta},\boldsymbol{m})$ conditional
on some observed $\boldsymbol{x}$, whereas Proposition \ref{prop:convergence}
in the current paper considers the distribution of $(\boldsymbol{\theta},\boldsymbol{m})$
unconditional on $\boldsymbol{x}$. With this adjustment, the same
argument in \citet{wei2024estimating} applies directly, so we omit
repeating the details here.
\end{document}